\documentclass[12pt]{article}
\pdfoutput=1 
\usepackage[margin=1in]{geometry}
\usepackage{graphicx}
\usepackage{tabularx}
\usepackage{multirow}
\usepackage{url}
\usepackage{amsmath, amsthm, amssymb, amsfonts}
\usepackage{mathtools}
\usepackage[usenames,dvipsnames]{xcolor}
\usepackage[colorlinks=true,linkcolor=Blue,citecolor=Blue,urlcolor=Blue]{hyperref}

\usepackage{authblk}
\usepackage[capitalise]{cleveref}
\creflabelformat{equation}{#2\textup{#1}#3}
\usepackage{xfp}
\usepackage{amsmath}

\title{A minimal model of peripheral clocks reveals differential circadian re-entrainment in aging}
\author[1,2]{Yitong Huang\footnote{Correspondence: yitong.huang@northwestern.edu}}
\author[3]{Yuanzhao Zhang}
\author[1,2,4,5,6]{Rosemary Braun\footnote{Correspondence: rbraun@northwestern.edu}}
\affil[1]{Department of Molecular Biosciences, Northwestern University, Evanston, IL, USA}
\affil[2]{NSF-Simons Center for Quantitative Biology, Northwestern University, Evanston, IL, USA}
\affil[3]{Santa Fe Institute, Santa Fe, NM, USA}
\affil[4]{Department of Engineering Sciences and Applied Mathematics, Northwestern University, Evanston, IL, USA}
\affil[5]{Department of Physics and Astronomy, Northwestern University, Evanston, IL, USA}
\affil[6]{Northwestern Institute on Complex Systems, Northwestern University, Evanston, IL, USA}

\begin{document}
\maketitle
\begin{abstract}
The mammalian circadian system comprises a network of cell-autonomous oscillators, spanning from the central clock in the brain to peripheral clocks in other organs. These clocks are tightly coordinated to orchestrate rhythmic physiological and behavioral functions. Dysregulation of these rhythms is a hallmark of aging, yet it remains unclear how age-related changes lead to more easily disrupted circadian rhythms. Using a two-population model of coupled oscillators that integrates the central clock and the peripheral clocks, we derive simple mean-field equations that can capture many aspects of the rich behavior found in the mammalian circadian system. We focus on three age-associated effects which have been posited to contribute to circadian misalignment: attenuated input from the sympathetic pathway, reduced responsiveness to light, and a decline in the expression of neurotransmitters. We find that the first two factors can significantly impede re-entrainment of the clocks following a perturbation, while a weaker coupling within the central clock does not affect the recovery rate. Moreover, using our minimal model, we demonstrate the potential of using the feed-fast cycle as an effective intervention to accelerate circadian re-entrainment. These results highlight the importance of peripheral clocks in regulating the circadian rhythm and provide fresh insights into the complex interplay between aging and the resilience of the circadian system.
\end{abstract}

\section{Introduction}
Circadian clocks are cell-autonomous oscillators that regulate physiological and behavioral processes with a 24~hour period. The mammalian circadian system consists of a hierarchical network of oscillators, where the central clock in the suprachiasmatic nucleus (SCN) is thought to orchestrate the peripheral clocks found in nearly every tissue of the body~\cite{dibner2010mammalian,honma2018mammalian}.  Additionally, each population of clocks may entrain to different environmental cues (zeitgebers), such as light, temperature, feeding, or activity~\cite{stratmann2006properties,brown2013peripheral}.  While the central clock primarily responds to the light-dark cycle~\cite{dibner2010mammalian,honma2018mammalian} and the peripheral clocks respond to the feed-fast cycle~\cite{stratmann2006properties,brown2013peripheral}, the SCN also acts as a conductor, sending signals to synchronize the peripheral clocks through the secretion of melatonin. This raises the question of how the circadian system responds to different zeitgebers, particularly when they are discrepant. 

Circadian misalignment is commonly observed between the circadian system and the external environment, an experience shared by jet-lagged travelers and shift workers. Abrupt changes to one's daily schedule disrupt the timing of the circadian clock.  Chronic circadian misalignment (such as that induced by social jet lag and shift work) has been linked to adverse health outcomes, including cardiovascular disease, obesity, and cancer~\cite{costa1996impact,rajaratnam2001health,smith2012shift,boivin2022disturbance}.  Yet while most attention has been focused on the misalignment between the central SCN clock and the environment, desynchrony can also occur \textit{within} the circadian system, between the central clock and the peripheral tissues. For instance, in humans, eating at night activates the liver clock at a time when the SCN wants to rest. In a simulated night shift study, researchers observed that nighttime eating can lead to a misalignment between the endogenous central clock (core body temperature) and peripheral clocks (glucose)~\cite{chellappa2021daytime}.

It is well known that the process of aging is associated with changes in sleep and circadian functions. On the physiological level, circadian output from the SCN has been reported to decrease with age~\cite{hofman2006living,nakamura2016suprachiasmatic}, though the molecular clocks in the aged SCN and in peripheral tissues express normal oscillations in vivo~\cite{nakamura2016suprachiasmatic,nakamura2011age,yamazaki2002effects}. Previous work has also shown dampened rhythms in neuropeptides in the SCN and decreased glucose metabolism~\cite{buijink2021multi,defronzo1981glucose,hofman2000human}. In addition, the aging of the eyes also leads to less light reaching the SCN~\cite{charman2003age}. On the behavioral level, multiple studies in older humans and aged animal models demonstrate fragmented activity rhythms and a longer latency to re-entrain under simulated jet lag protocols~\cite{sellix2012aging,klerman2001circadian,monk2005aging,duffy2015aging}. Together, these findings suggest that aging might result in a circadian system that is more vulnerable to disruption and misalignment. Yet despite the consistent observations of age-related decline in the circadian system, 
we still lack a systematic understanding of the complex interplay between aging and circadian misalignment. For example, what role do peripheral clocks play in aging-induced circadian disruption?
Are there simple and practical interventions that we can take to alleviate such disruption?

To investigate age-related circadian misalignment, we first develop a mathematical model to describe the circadian clocks in the SCN and peripheral tissues. 
Over the years, there have been a plethora of models that describe circadian clocks with different levels of granularity, from complex mechanistic models that take into account the detailed dynamics of specific molecular interactions involving circadian genes and proteins~\cite{leloup2003toward,becker2004modeling,bernard2007synchronization,to2007molecular,vasalou2010multiscale,hafner2012effect} to more abstract models with node dynamics that aim to capture the collective behavior of circadian cells across the SCN~\cite{garcia2004modeling,gonze2005spontaneous,fukuda2007synchronization,ullner2009noise,komin2011synchronization,gu2015noise,gu2016heterogeneity,hannay2018macroscopic,hannay2019macroscopic,hannay2020seasonality,zhang2020energy,zhou2022network}.  In general, most of these studies have focused on the central circadian rhythm, and thus consider a single population of clocks. However, experimental data clearly demonstrate a hierarchical network of clocks in mammalian animals~\cite{dibner2010mammalian,honma2018mammalian}, and these clocks respond to different external stimuli~\cite{stratmann2006properties,brown2013peripheral}. Therefore, a mathematical framework containing both the central clock and peripheral clocks is necessary to advance our knowledge of circadian rhythms.
Here, we study the interplay between the SCN and peripheral clocks using a novel model based on two coupled populations of Kuramoto oscillators~\cite{strogatz2000kuramoto} and study how they integrate disparate zeitgebers.
Our model has a number of appealing properties: its simplicity allows us to perform semi-analytical analyses of the model through the Ott-Antonsen ansatz~\cite{ott2008low,ott2009long,lee2009large,kotwal2017connecting,engelbrecht2020ott}, yet it is sophisticated enough to capture many intriguing features of the real circadian clocks, such as the east-west asymmetry of circadian response to jet lag~\cite{lu2016resynchronization,song2017jet}.  

We then use this model to investigate the effect of aging by focusing on three age-associated changes that have been conjectured to contribute to circadian misalignment:
 (i)~attenuated sympathetic nervous system from the SCN~\cite{nakamura2011age,tahara2017age}; 
  (ii)~reduced sensitivity to light~\cite{charman2003age}; 
  and (iii)~reduced coupling strengths in the SCN~\cite{nakamura2016suprachiasmatic,de1982neuropeptides}. 
We examine how each condition alters the re-entrainment pattern in a 6-hr phase advance and a 6-hr phase delay protocol. 
Based on these results, we also propose a simple intervention targeting the peripheral clocks that can potentially alleviate circadian misalignment. 

Our investigations make four high-level contributions. First, we propose a novel multi-oscillator framework to integrate the central clock and peripheral clocks with two different entraining signals. Second, we find that both attenuated sympathetic pathway signals and reduced sensitivity to light can lead to longer recovery time from abrupt shifts in the entraining signal, which also exhibits asymmetry towards phase delay and phase advance. 
Moreover, an attenuated sympathetic pathway affects the peripheral clocks, whereas a reduced sensitivity mostly disrupts the central clock. Third, we find that reduced coupling strength in the central clock is responsible for its reduced output amplitude, but it does not alter the re-entrainment process. Finally, we highlight the importance of food stimuli and peripheral clocks: a properly adjusted meal schedule can significantly speed up the re-entrainment process after circadian disruption such as jet lag.

\section{Method: deriving a minimal model of the central and peripheral clocks}
\begin{figure}[t]
    \centering
    \includegraphics[width = .7\textwidth]{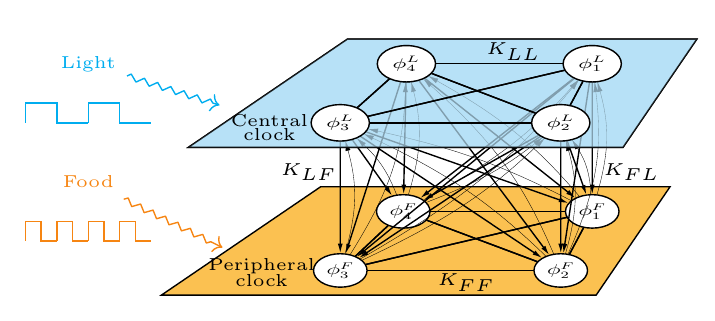}
    \caption{Schematic of the mathematical model. The model consists of two populations of coupled oscillators, where one population represents the central clock in the brain, entrained by light, and the other population represents the peripheral clocks, entrained by food. We assume that all oscillators are coupled to one another, but the intralayer and interlayer coupling strengths can be different. Moreover, the downward coupling (from the central clock to the peripheral clocks) is stronger than the upward coupling.}
    \label{fig:scheme}
\end{figure}

In our model (\cref{fig:scheme}), we consider two populations of cells described by their internal phase $\phi_k^F$ and $\phi_k^L$---one food-entrained (e.g., liver cells) and the other light-entrained (e.g., SCN cells):
\begin{align}
    \frac{d\phi_k^F}{dt} &= \omega_k^F + \underbrace{\frac{K_{FF}}{N_F}\sum_{j=1}^{N_F}\sin(\phi_j^F - \phi_k^F)}_\text{intralayer coupling} + \underbrace{\frac{K_{LF}}{N_L}\sum_{j=1}^{N_L}\sin(\phi_j^L - \phi_k^F + \alpha)}_\text{interlayer coupling} + \underbrace{F(t)M(\phi_k^F)
    \vphantom{\sum_{j=1}^{N_F}}
    }_\text{food stimulus}\,, \label{eq:model1} \\[1em]
    \frac{d\phi_k^L}{dt} &= \omega_k^L + \underbrace{\frac{K_{LL}}{N_L}\sum_{j=1}^{N_L}\sin(\phi_j^L - \phi_k^L)}_\text{intralayer coupling} + \underbrace{\frac{K_{FL}}{N_F}\sum_{j=1}^{N_F}\sin(\phi_j^F - \phi_k^L - \alpha)}_\text{interlayer coupling} + \underbrace{L(t)Q(\phi_k^L)
    \vphantom{\sum_{j=1}^{N_F}}
    }_\text{light stimulus}\,. \label{eq:model2}
\end{align}
%
Here, $\omega_k^F$ ($\omega_k^L$) is the natural frequency of the $k$-th food-entrained (light-entrained) oscillator, which is drawn from a distribution $g(\omega)$; $\alpha$ is the homeostatic (physiologically desirable) phase difference between the populations; $F(t)$ and $L(t)$ denote the stimulus of food and light respectively; $M(\phi)$ gives the phase response curve of the food-entrained oscillators, and $Q(\phi)$ gives the phase response curve of the light-entrained oscillators.
The phase response curves describe the change in phase of the various oscillators as a function of the time at which the zeitgeber stimulus is received.
Here we assume that the food and light phase response curves have the same shape, but potentially different sensitivity. 
That is, $Q(\phi) = cM(\phi)$, where $c$ is a constant sensitivity ratio. 
The coupling strengths are given as $K_{\text{from},\text{to}}$ (e.g., $K_{LF}$ is the coupling strength from light-entrained oscillators to food-entrained oscillators), which are normalized by the number of oscillators in the corresponding layer. 
Below, wherever appropriate, we will suppress the population indicator ($F$ and $L$) to ease the notation.

In the continuum limit ($N\rightarrow\infty$), each oscillator population can be described by the phase density function $f(\omega, \phi, t)$, which gives the density of oscillators with phase $\phi$ at time $t$ and have frequency $\omega$. The time-evolution of $f$ is governed by a continuity equation (which basically states that the total density or the number of oscillators is conserved):
\begin{equation}
    \frac{\partial f}{\partial t} + \frac{\partial}{\partial \phi}(fv) = 0\,,
    \label{eq:continuity}
\end{equation}
where $v = \dot{\phi}$.
Using the integral form of \cref{eq:model1,eq:model2} in the continuum limit, we can express $v_F = \dot{\phi}_F$ as
\begin{multline}
    v_F(\omega, \phi, t)=\omega + K_{FF} \int_{-\infty}^{\infty} \int_{-\pi}^\pi \sin \left(\phi^{\prime}-\phi\right) f_F\left(\omega^{\prime}, \phi^{\prime}, t\right) d \omega^{\prime} d \phi^{\prime} \\
    + K_{LF} \int_{-\infty}^{\infty} \int_{-\pi}^\pi \sin \left(\phi^{\prime}-\phi+\alpha\right) f_L\left(\omega^{\prime}, \phi^{\prime}, t\right) d \omega^{\prime} d \phi^{\prime} + F(t)M(\phi)\,,
    \label{eq:v_F}
\end{multline}
An analogous equation can be obtained for $v_L = \dot{\phi_L}$.

For a large population of coupled phase oscillators, it is useful to define the Daido order parameters of the phase distribution
\begin{equation}
    Z_m(t) = R_m(t) e^{i\psi_m(t)} =\frac{1}{N}\sum_{j=1}^{N} e^{im\phi_j(t)}\,,
    \label{eq:z_m_finite}
\end{equation}
where 
$R_m$ are the collective amplitudes that measure phase coherence and $\psi_m$ are the mean phases. 
The first term, the Kuramoto order parameter $Z_1$,
measures the collective amplitude, with 
$R_1=0$ indicating complete desynchrony and 
$R_1=1$ indicating perfectly synchronized oscillators. 
In the limit $N\rightarrow\infty$, \cref{eq:z_m_finite} can be written as
\begin{equation}
    Z_m(t) = \int_0^{2\pi}\int_{-\infty}^{\infty} f(\omega,\phi,t)e^{i m \phi} \; d\omega \; d\phi\,.
    \label{eq:z_m_infinite}
\end{equation}
This allows us to write \cref{eq:v_F} in a more compact form using the Daido order parameters:
\begin{multline}
    v_F(\omega, \phi, t)=\omega + \frac{1}{2i} K_{FF} \left( Z_1^F e^{-i\phi} - \overline{Z_1^F} e^{i\phi} \right) \\
    + \frac{1}{2i} K_{LF} \left( Z_1^L e^{-i(\phi-\alpha)} - \overline{Z_1^L} e^{i(\phi-\alpha)} \right) + F(t)M(\phi)\,.
    \label{eq:v_F_simple}
\end{multline}

Next, we consider the Fourier expansion of the phase density function $f$,
\begin{equation}
    f = \frac{g(\omega)}{2\pi}\left(1+\sum_{k=1}^\infty a_k(\omega,t)e^{ik\phi}+ \text{c.c.}\right)\,,
    \label{eq:fourier}
\end{equation}
where c.c.\ stands for the complex conjugate of the expression and $g(\omega)$ is the distribution of natural frequencies of the oscillators.
Moreover, for simplicity, we assume the phase response curves are given in the following form:
\begin{align}
    M(\phi) &= \sigma_F - A_1\sin(\phi + \zeta_1) - A_2\sin(2\phi+\zeta_2)\,,\\
    Q(\phi) &= \sigma_L - B_1\sin(\phi + \beta_1) - B_2\sin(2\phi+\beta_2)\,,
\end{align}
where the parameters $\sigma_F, \sigma_L, A_1, A_2, B_1, B_2, \zeta_1, \zeta_2, \beta_1,$ and $\beta_2$ are coefficients of the truncated harmonic expansion for $M(\phi)$ and $Q(\phi)$, respectively. This truncated harmonic expansion is motivated by empirical measurements and can be fitted to available experimental data~\cite{hannay2018macroscopic,hannay2019macroscopic}.
We also note that the derivation below can be easily adapted to other forms of phase response curves.


Now, substituting the Fourier expansion of $f$ (\cref{eq:fourier}) and the expression for $v$ (\cref{eq:v_F_simple}) into the continuity equation (\cref{eq:continuity}), then collecting the coefficients in front of $e^{ik\phi}$ for each $k$, we arrive at a set of infinitely many coupled integral-differential equations for the Fourier coefficients $\{a_k^F\}$ and $\{a_k^L\}$:
\begin{multline}
    \frac{\dot{a}_k^F}{k} + \big(i\omega +i\sigma_F F(t)\big)a_k^F + \frac{K_{FF}}{2}\Big(Z_1^F a_{k+1}^F - \overline{Z_1^F} a_{k-1}^F\Big) +\frac{K_{LF}}{2}\Big(Z_1^L a_{k+1}^F e^{i\alpha} - \overline{Z_1^L} a_{k-1}^F e^{-i\alpha}\Big)\\ +\frac{1}{2}\Big(H_1^F a_{k+1}^F - \overline{H_1^F} a_{k-1}^F \Big) +\frac{1}{2}\Big(H_2^F a_{k+2}^F - \overline{H_2^F} a_{k-2}^F\Big) = 0\,,
    \label{eq:a_k^F}
\end{multline}
\begin{multline}
    \frac{\dot{a}_k^L}{k} + \big(i\omega +i\sigma_L L(t)\big)a_k^L + \frac{K_{LL}}{2}\Big(Z_1^L a_{k+1}^L - \overline{Z_1^L} a_{k-1}^L\Big) +\frac{K_{FL}}{2}\Big(Z_1^F a_{k+1}^L e^{-i\alpha} - \overline{Z_1^F} a_{k-1}^L e^{i\alpha}\Big)\\ +\frac{1}{2}\Big(H_1^L a_{k+1}^L - \overline{H_1^L} a_{k-1}^L \Big) +\frac{1}{2}\Big(H_2^L a_{k+2}^L - \overline{H_2^L} a_{k-2}^L\Big) = 0\,,
    \label{eq:a_k^L}
\end{multline}
where
\begin{align*}
H_1^F(t) = A_1 e^{-i\zeta_1} F(t), & \quad H_2^F(t) = A_2 e^{-i\zeta_2} F(t),\\[1ex]
H_1^L(t) = B_1 e^{-i\beta_1} L(t), & \quad H_2^L(t) = B_2 e^{-i\beta_2} L(t).
\end{align*}

To make further progress, we now make the standard assumption that the intrinsic frequencies of each oscillator population follow a Cauchy (Lorentzian) distribution 
\begin{equation*}
    g(\omega) = \frac{1}{\pi}\frac{\gamma}{(\omega -\omega_0)^2+{\gamma}^2} 
\end{equation*}
with mean frequency $\omega_0$ and dispersion parameter ${\gamma}$.
Due to the orthogonality of the Fourier basis functions, in the continuum limit, the Daido order parameters $Z_m$ can be given by 
\begin{equation}
    Z_m(t) = \int_0^{2\pi}\int_{-\infty}^{\infty} f(\omega,\phi,t)e^{i m \phi} \; d\omega \; d\phi = \int_{-\infty}^{\infty} \overline{a_m}(\omega,t)g(\omega) \; d\omega\,.
    \label{eq:z_m}
\end{equation}
The Cauchy distribution assumed for $g(\omega)$ enables us to evaluate \cref{eq:z_m} analytically through contour integration, simplifying $Z_m$ to 
\begin{equation}
    Z_m(t) = \overline{a_m}(\omega_0-i\gamma, t)\,.
    \label{eq:Daido}
\end{equation}
Using \cref{eq:Daido} and setting $\omega=\omega_0-i\gamma$ in \cref{eq:a_k^F,eq:a_k^L}, we can replace the Fourier coefficients with Daido order parameters, yielding
\begin{multline}
    \frac{\dot{Z_k^F}}{k}  = \left(-\gamma_F + i\omega_0^F + i\sigma_F F(t)\right)Z_k^F + \frac{K_{FF}}{2}\Big(Z_1^F Z_{k-1}^F - \overline{Z_1^F} Z_{k+1}^F\Big) \\+ \frac{K_{LF}}{2}\Big(Z_1^L Z_{k-1}^F e^{i\alpha} - \overline{Z_1^L}Z_{k+1}^F e^{-i\alpha}\Big) +\frac{1}{2}\Big(H_1^F Z_{k-1}^F - \overline{H_1^F}Z_{k+1}^F\Big) \\+\frac{1}{2}\Big(H_2^F Z_{k-2}^F -\overline{H_2^F}Z_{k+2}^F\Big)\,,
\end{multline}
\begin{multline}
    \frac{\dot{Z_k^L}}{k}  = \left(-\gamma_L + i\omega_0^L + i\sigma_L L(t)\right)Z_k^L + \frac{K_{LL}}{2}\Big(Z_1^L Z_{k-1}^L - \overline{Z_1^L} Z_{k+1}^L\Big) \\+ \frac{K_{FL}}{2}\Big(Z_1^F Z_{k-1}^L e^{-i\alpha} - \overline{Z_1^F}Z_{k+1}^L e^{i\alpha}\Big) +\frac{1}{2}\Big(H_1^L Z_{k-1}^L - \overline{H_1^L}Z_{k+1}^L\Big) \\+\frac{1}{2}\Big(H_2^L Z_{k-2}^L -\overline{H_2^L}Z_{k+2}^L\Big)\,.
\end{multline}
In particular, $k=1$ represents the equations of motion for the Kuramoto order parameters $Z_1^F$ and $Z_1^L$.

However, these equations are not closed due to the presence of $Z_2^F$ and $Z_3^F$ ($Z_2^L$ and $Z_3^L$) on the right-hand side.
To close the equations for $Z_1^F$ and $Z_1^L$, we invoke the Ott-Antonsen ansatz, $Z_m = Z_1^m$ (or equivalently, $R_m = R_1^m$ and $\psi_m = m\psi_1$), which forms a low-dimensional flow-invariant manifold under \cref{eq:continuity} and has been shown to be globally attracting under weak conditions~\cite{ott2009long,engelbrecht2020ott}.
Applying the ansatz and 
rewriting the equations in polar coordinates
yields a four-dimensional system describing the phase coherence $R_{F,L}$ and mean phase $\psi_{F,L}$ of each cluster: 
\begin{align}
    \dot{R_F} = & -\gamma_F R_F + \frac{K_{FF}}{2}(R_F- R_F^3) +\frac{K_{LF}}{2}R_L(1-R_F^2)\cos(\psi_L-\psi_F+\alpha) + U_F(R_F,\psi_F)\,, \\[1ex]
    \dot{\psi_F} = & \omega_0^F +\sigma_F F(t) +\frac{K_{LF}}{2}R_L\left(\frac{1}{R_F}+R_F\right)\sin(\psi_L-\psi_F+\alpha)+V_F(R_F,\psi_F)\,,\\[1ex]
    \dot{R_L} = & -\gamma_L R_L + \frac{K_{LL}}{2}(R_L- R_L^3) +\frac{K_{FL}}{2}R_F(1-R_L^2)\cos(\psi_F-\psi_L-\alpha) + U_L(R_L,\psi_L)\,, \\[1ex]
    \dot{\psi_L} = & \omega_0^L +\sigma_L L(t) +\frac{K_{FL}}{2}R_F\left(\frac{1}{R_L}+R_L\right)\sin(\psi_F-\psi_L-\alpha)+V_L(R_L,\psi_L)\,,
\end{align}
where the stimulus terms are given by
\begin{align}    
    U_F(R_F,\psi_F) & = \frac{A_1}{2}F(t)(1-R_F^2)\cos(\zeta_1 +\psi_F) +\frac{A_2}{2}F(t)R_F(1-R_F^2)\cos(\zeta_2+2\psi_F)\,,\\[1ex]
    V_F(R_F,\psi_F) & = -\frac{A_1}{2}F(t)\left(\frac{1}{R_F}+R_F\right)\sin(\zeta_1 +\psi_F) -\frac{A_2}{2}F(t)(1+R_F^2)\sin(\zeta_2+2\psi_F)\,,\\[1ex]
    U_L(R_L,\psi_L) & = \frac{B_1}{2}L(t)(1-R_L^2)\cos(\beta_1 +\psi_L) +\frac{B_2}{2}L(t)R_L(1-R_L^2)\cos(\beta_2+2\psi_L)\,,\\[1ex]
    V_L(R_L,\psi_L) & = -\frac{B_1}{2}L(t)\left(\frac{1}{R_L}+R_L\right)\sin(\beta_1 +\psi_L) -\frac{B_2}{2}L(t)(1+R_L^2)\sin(\beta_2+2\psi_L)\,.
\end{align}

\section{Results: numerical simulations of re-entrainment dynamics}
Several studies in both human and animal models have demonstrated that aging results in a longer latency to re-entrain following abrupt phase shifts. A primary goal of our study is to unravel the complex interplay between aging, the central clock, and peripheral clocks in the re-entrainment process. We consider the re-entrainment process is complete when the following two criteria are met: 
\begin{enumerate}
    \item the collective phase of each clock becomes stable, that is, the fluctuations in each clock do not exceed $0.05$ radians (about $2.86^{\circ}$ or 0.2~hours) between consecutive days;
    \item the difference between the collective phase of the central clock and the collective phase of the peripheral clocks returns to approximately the same after the shift. That is, the fluctuations in the phase gap of two clocks do not exceed $0.05$ radians before and after the shift. 
\end{enumerate}

To facilitate our analysis, we start with a default parameter set for the model to represent a typical young and healthy person. Since previous studies have developed and validated the physiological model of the SCN, the parameters of the central clock ($\gamma_L, \omega_L, K_{LL}, Q(\phi)$) in our model are chosen according to published studies~\cite{hannay2018macroscopic,hannay2019macroscopic}. However, less is known about the peripheral clocks and, in particular, the phase response curve to food has not been established. Here, we choose the parameters of the peripheral clocks according to the following observations and assumptions: 
\begin{enumerate}
    \item peripheral clocks express self-sustained 24-hr oscillations~\cite{yoo2004period2}; 
    \item there is a phase lag of $\sim4$ hr for the coupling between the central clock and peripheral clocks (since the peaks of mPER2 rhythms in the SCN and peripheral tissues differ from 3 hr to 9 hr in mice~\cite{yoo2004period2});
    \item the downward coupling from the brain to the peripheral organs is stronger than the upward coupling because of the hierarchical nature of the circadian system and the brain-blood barrier;
    \item the shape of the phase response curve to food in peripheral clocks is the same as the phase response to light in the central clock.
\end{enumerate}
Based on these assumptions, we choose our default parameter set as shown in \cref{table1}. Later, we will vary certain parameters to mimic different aspects of aging-induced change in circadian physiology.

\begin{table}[h!]
\centering
\begin{tabular}{cc} 
\hline
Parameters & Value \\
\hline
$K_{LL}$, $K_{FF}$ & 0.065 \\ 
$K_{LF}$ & 0.06 \\ 
$K_{FL}$ & 0.0005 \\ 
$\gamma_F$& 0.03 \\ 
$\gamma_L$& 0.024 \\ 
$\alpha$& 1 \\ 
$A_1, B_1$& 0.4 \\ 
$A_2, B_2$& 0.2 \\ 
$\alpha_1, \beta_1$& 0.2 \\ 
$\alpha_2, \beta_2$& -1.8 \\ 
$\sigma_f, \sigma_f$& 0.05 \\ 
\hline
\end{tabular}
\caption{Default parameter set used in numerical simulations to model the circadian clock of a young and healthy adult.}
\label{table1}
\end{table}


Next, we assume the light-dark cycle of each day consists of 12 hours of light and 12 hours of darkness. Empirical evidence shows that the feed-fast cycle is more important than the light-dark cycle for circadian oscillations in peripheral clocks, so we minimize the effect of the feeding cues by providing scheduled feeding four times per day at equally spaced intervals, as in the constant routine experiments~\cite{chellappa2021daytime,tahara2017age}. Under this condition, we study the effect of an abrupt shift of light schedule (e.g., caused by traveling across time zones) on the circadian system. 

\subsection{Attenuated input from the sympathetic pathway}
\begin{figure}[h!]
    \centering
    \includegraphics[width = \textwidth]{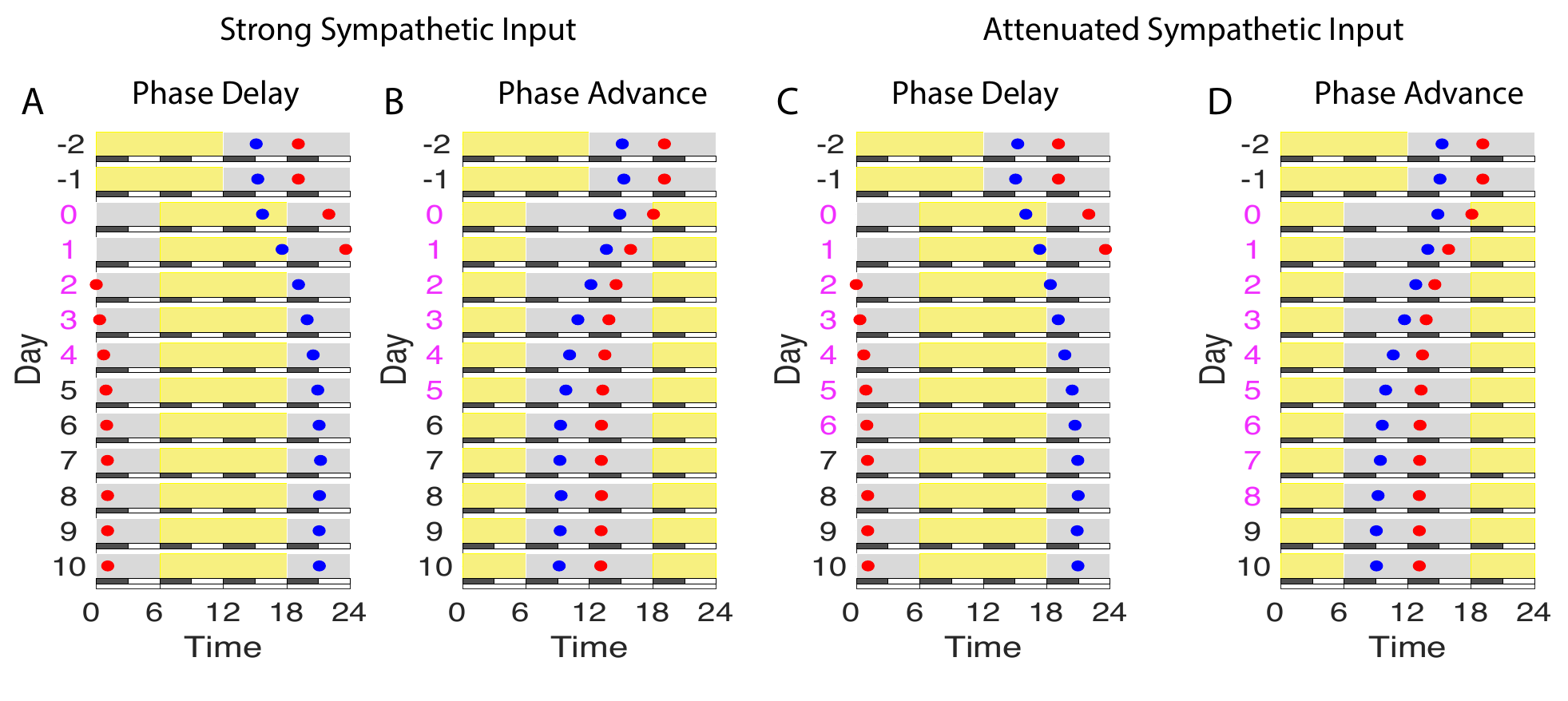}
    \caption{Effect of an attenuated sympathetic pathway on the re-entrainment process. Estimated phase markers of the peripheral clock (blue dot) and the central clock (red dot) are plotted against the pattern of the light-dark cycle and feed-fast cycle. The light-dark cycle is composed of 12 hours of light (yellow) and 12 hours of darkness (gray). The feed-fast cycle consists of 4 equally distributed meals a day, where the black bar represents the food availability. The subjects are initially entrained to a 12:12 light:dark cycle. On day 0, the schedule shift occurs. Two schedules, a 6-hr phase delay, and a 6-hr phase advance are compared. Times during re-entrainment are colored in magenta. (A) and (B) represent the re-entrainment process for a young subject using the default parameter set. We can see that both the central clock and the peripheral clock in a young subject complete the phase-delay re-entrainment in 5 days and complete the phase-advance entrainment in 6 days. To model an age-associated decline in the sympathetic nervous system, we reduce the strength of the downward coupling $K_{LF}$ from 0.06 (default parameter set, A and B) to 0.035 (C and D). With a reduced value of $K_{LF}$, the central clock (red dots) is not affected, but the peripheral clock (blue dots) exhibits slower adjustment compared to (A) and (B). As a result, an attenuated sympathetic pathway leads to longer re-entrainment processes---7 days for phase delay and 9 days for phase advance.}
    \label{fig2}
\end{figure}

In \cref{fig2}, we demonstrate the re-entrainment process to a 6-hour shift of the light-dark cycle with the potential age-associated effect on the sympathetic pathway. Briefly, we simulate the above system with the baseline light:dark cycle and regular feeding for two days. We then abruptly shift the light:dark cycle by 6~hours and continue the simulation to examine the phase of the central and peripheral clocks ($\psi_L$ and $\psi_F$) as they re-entrain to the new conditions.  In the baseline entrainment, days $-2$ to $-1$ show a light-dark schedule of 12~hours of light and 12~hours of darkness. At day 0, we assume a transition occurs (e.g., going from Paris to New York for \cref{fig2}A and C; from New York to Paris for \cref{fig2}B and D). In \cref{fig2}, each new line represents a subsequent day. Gray denotes darkness, and yellow denotes light.  
As a phase marker, we choose the time at which $\psi=\pi$ to represent the phase of each clock. 
%
A blue dot shows the predicted phase of the peripheral clocks, and a red dot shows the predicted timing of the central clock. Times during re-entrainment are marked in magenta. We consider the system fully re-entrained to the new light schedule when the fluctuations in both the individual clock and the phase gap between clocks do not exceed 0.05 (in radians) between consecutive days. 

Empirical evidence suggests that peripheral clock input from the sympathetic nervous system is attenuated as aging progresses~\cite{nakamura2011age,tahara2017age}. The strength of this input can be modeled in our framework by the coupling term $K_{LF}$. 
\cref{fig2}(A) and \cref{fig2}(B) show the re-entrainment process when the downward coupling is strong ($K_{LF}=0.06$), as it might be for young healthy adults. 
We can observe that re-entrainment takes 5~days for a westward trip (6~hr phase delay) and 6~days for an eastward trip (6~hr phase advance).  This asymmetry is due to asymmetries in the phase-response curve, and is consistent with observations that westward travel is generally easier than eastward~\cite{lu2016resynchronization,song2017jet,monk2000inducing}.
Thus, for a young adult represented by the default parameter set, re-entrainment time to 6-hr phase advance and 6-hr phase delay  differ by one day. 
To investigate the effect of a weaker sympathetic pathway, we reduce $K_{LF}$ from $0.06$ to $0.035$. This results in a less synchronized state in the peripheral tissues and a slower adjustment of the peripheral clock at the beginning of the transition. 
(As expected, the central clock is unaffected by the change in $K_{LF}$, see \cref{fig:sup1}.) 
This change prolongs the overall recovery (time for both clocks to be synchronized to the new environment) but does so asymmetrically.  We find that lower $K_{LF}$ adds two days to the overall recovery time in the 6~hour phase delay condition (that is, a total of 7 days for both clocks to be synchronized again). 
On the other hand, the same reduced value of $K_{LF}$ prolongs the recovery time for the 6~hr phase advance by three days (to a total of 9 days. This suggests that reduce sympathetic coupling may account for the increasing difficulty of jetlag recovery as people age.  It also suggests that the asymmetry of phase adjustments may worsen with age, where phase advance (eastward travel) has a much slower recovery time than a phase delay (westward travel). 

\subsection{Decreased photic resetting}
\begin{figure}[ht]
    \centering
    \includegraphics[width = 0.5\textwidth]{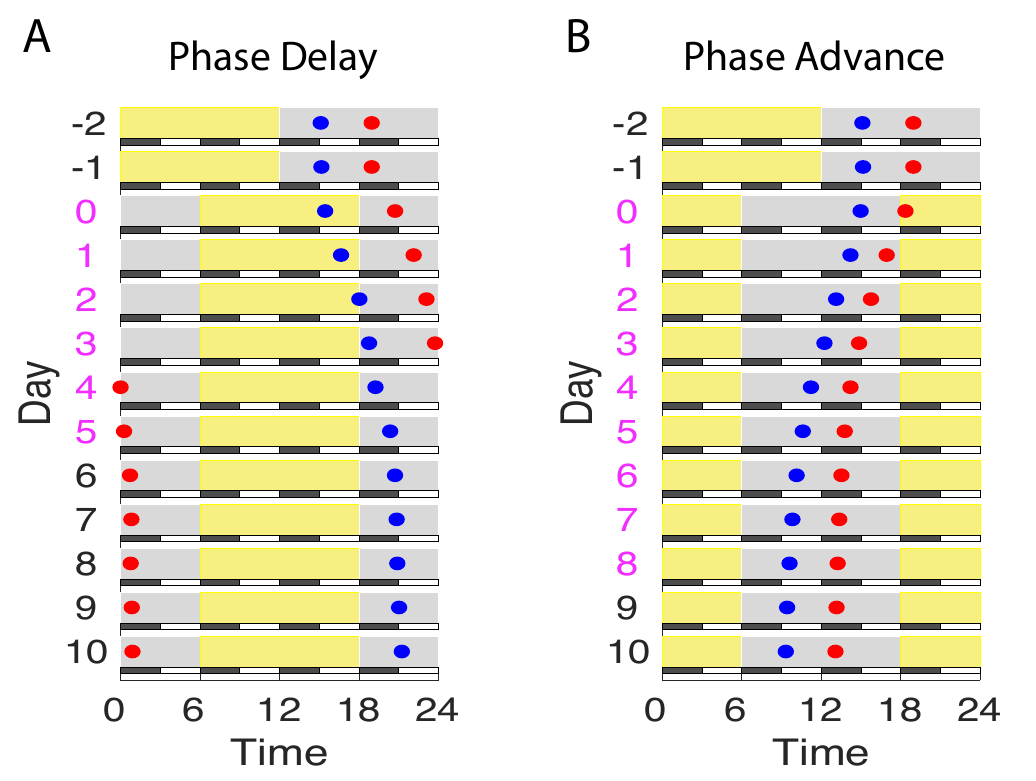}
    \caption{Effect of reduced sensitivity to light. As in \cref{fig2}, two phase markers of the central clock (red dot) and the peripheral clock (blue dot) are presented and used to measure the re-entrainment process. Times during re-entrainment are colored in magenta. Here, we apply the default parameter set but scale the light response by 0.6 to represent an age-associated reduced response to light. We find that a weaker response to light results in a slower re-entrainment rate of the central clock, represented by the red dots. (A) shows that the phase-delay re-entrainment completes in 6 days, and (B) shows that it needs 9 days to complete a phase-advance. }
    \label{fig3}
\end{figure}

Changes with age in the cornea, lens, and pupils of the eyes have been well established~\cite{charman2003age}. As a result, the strength of light information flow to the central clock tends to decline as age advances. To examine the change in photic resetting resulting from these changes, we scale the effect of light stimuli $Q(\phi)$ by a constant fraction ($c=0.6$) to capture a lower sensitivity of the response to light. In \cref{fig3}, we show the recovery time with a decreased photic sensitivity for a 6-hr phase delay and 6-hr phase advance with other parameters at their default values. According to \cref{eq:model2}, the parameter $Q(\phi)$ contributes to the degree of synchronization of the central clock. Indeed, a decreased amplitude of $Q(\phi)$ leads to a less synchronized state of the central clock (\cref{fig:sup2}). From \cref{fig3}, we can see that the central clock, which responds to light, displays a slower rate of re-entrainment to a new schedule for a decreased strength of $Q(\phi)$. Surprisingly, we find that with a weaker sensitivity to light, both clocks can complete phase delay in 5-6 days (\cref{fig3}A), which is similar to the time required for a young subject with a normal light sensitivity (\cref{fig2}A). However, the recovery time to return to a phase-locked state for two clocks significantly increases for the phase-advance schedule. Having a weaker sensitivity to light extends the phase-advance re-entrainment from 6 days to 9 days. This result has two important consequences. First, it explains previous observations in clinical studies for the circadian response to light in aging, which found similar phase-adjustment patterns for old adults relative to young adults in the phase-delay direction, but reduced responses to light for older adults in phase advance~\cite{klerman2001circadian,benloucif2006responsiveness,duffy2007decreased}. Second, this result underscores the importance of strong light exposure in entrainment. 

\subsection{Reduced coupling within the central clock}
\begin{figure}[h!]
    \centering
    \includegraphics[width = 0.5\textwidth]{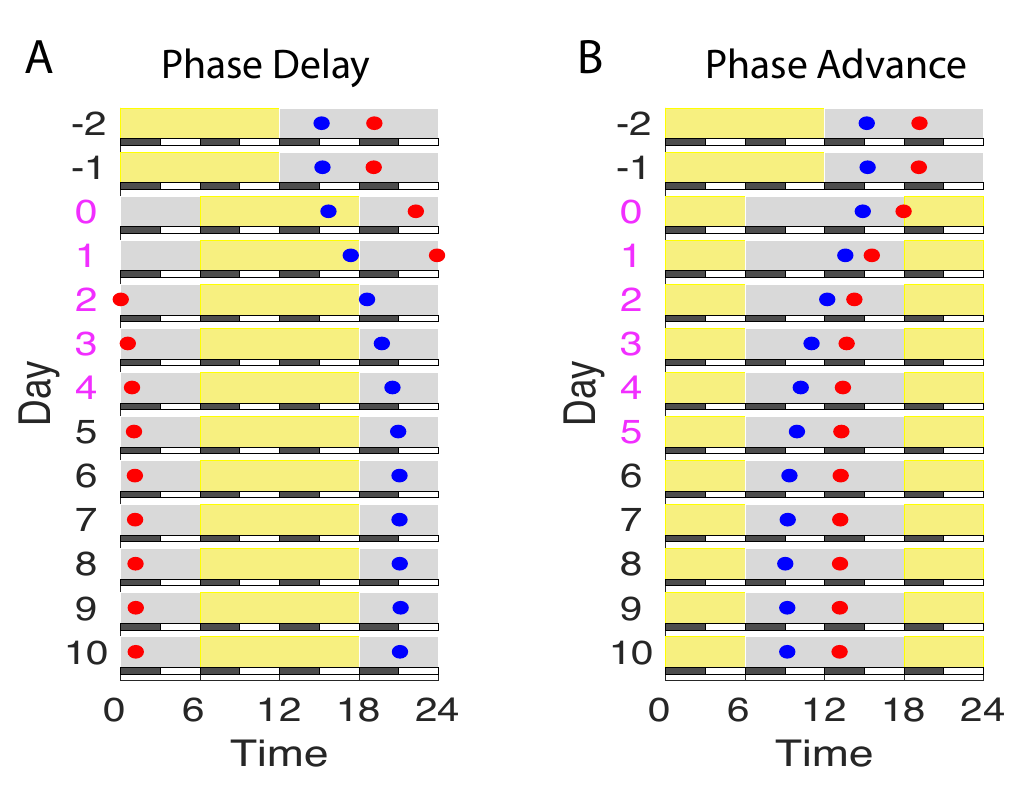}
    \caption{Effect of reduced coupling between the SCN cells. Here we reduce the intra-coupling term within the central clock, $K_{LL}$ from 0.065 to 0.03. The other parameters are kept the same as in the default parameter set. We can see that for both phase-delay (A) and phase-advance(B), a reduced coupling term in the central clock do not alter the recovery rate.
    }
    \label{fig4}
\end{figure}

Several studies have reported an age-related reduction in the amplitude of circadian rhythms, including gene expression, melatonin, and core body temperature rhythms~\cite{duffy2015aging,wolff2022defining}. However, the question remains of why such a reduction occurs. One possibility is that aging may cause reduced coupling strengths between cells within the clock. For instance, animal studies have found an age-related decline in neuropeptides in SCN cells~\cite{nakamura2016suprachiasmatic,de1982neuropeptides}. In our framework, the parameter $K_{LL}$, which represents the coupling strength within the central clock, controls the degree of synchronization in the central clock. Indeed, we observe a significantly decreased collective amplitude in the central clock as $K_{LL}$ decreases from 0.065 to 0.03 (\cref{fig:sup3}). However, we do not find that a decline in the intra-coupling strengths changes the rate of the re-entrainment process when external stimuli are present (\cref{fig4}). The recovery rate is the same as a young subject simulated using the default parameter set (\cref{fig2} A and B). In other words, our model suggests that reduced coupling between cells within the central clock is unlikely to contribute to the effect of aging on re-entrainment, but it might explain a dampened amplitude in circadian rhythms. 

\subsection{Adjusting meal schedule to accelerate re-entrainment}
\begin{figure}[h!]
    \centering
    \includegraphics[width = .5\textwidth]{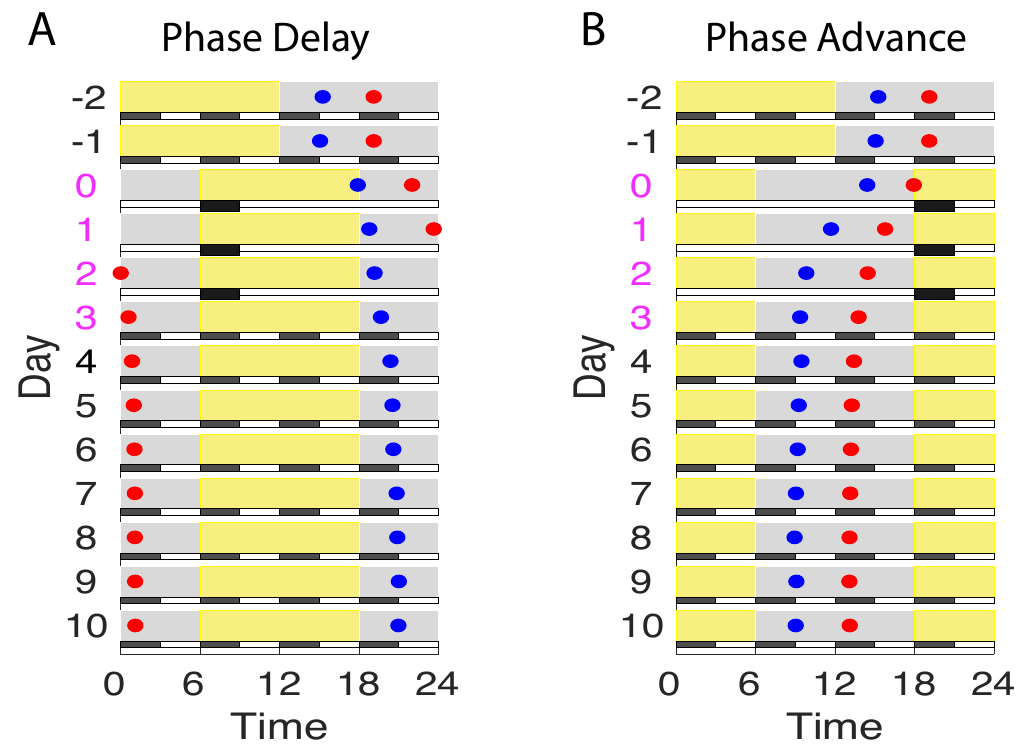}
    \caption{Meal interventions to accelerate the re-entrainment process. As in \cref{fig2}C and D, we consider an old subject with a weaker input from the sympathetic pathway ($K_{LF} = 0.035$), where it needed 7 days to phase delay and 9 days to phase advance. Instead of having uniformly distributed meals throughout the journey, the subject only has one meal a day in the early morning during the first three days of the transition, but this meal contains three times the volume. We can see that having such a meal schedule accelerates both phase-delay (A) and phase-advance(B) re-entrainment process to 4 days. }
    \label{fig5}
\end{figure}

Finally, we explore whether adjusting meals can help alleviate jet lag. From the previous section, we have seen that an age-related decline in the sympathetic pathway can affect the synchrony in the peripheral clocks and delay the re-entrainment process for both phase advance and phase delay. Therefore, to investigate the effect of meals on synchronization and re-entrainment, we set $K_{LF} = 0.035$ as in the previous section to represent an age-related decline that exists in the sympathetic nervous system. In addition, we simulate a different meal schedule in the first three days of re-entrainment and return to the equally distributed meal schedule afterwards. Instead of having uniformly distributed meals, only one meal containing three times the volume is given daily from day 0 to day 3. Compared to the recovery time in \cref{fig2}C and D, which requires 7 days for phase delay and 9 days for phase advance, \cref{fig5} shows that having a big meal in the early morning can accelerate the recovery time to 4 days for both a 6-hr phase delay and phase advance. As one can expect, having a meal at a different time, such as late at night, will result in an anti-phase relationship between the clocks. Moreover, such a meal schedule promotes synchrony in the peripheral clocks throughout the re-entrainment process (\cref{fig:sup4}). Therefore, our simulations highlight the importance of meal schedules and suggest the possibility of correcting the circadian misalignment by targeting peripheral clocks. 

\section{Discussion}
Daily rhythms of physiology and behavior were originally thought to be driven by the sleep{\slash}wake cycle.  The surprising discovery that nearly every cell possesses an autonomous circadian oscillator dramatically altered our understanding of circadian rhythms, and it is now understood that both the central (SCN) and peripheral clocks contribute to circadian organization~\cite{stratmann2006properties}.  Yet this discovery also opened  many additional questions: how do the cellular clocks remain synchronized, and how do they ``reset'' when given new time cues?

To study these questions, we developed a simple model comprising two populations of coupled oscillators that can describe the hierarchical nature of the circadian system. Our model significantly extends a previously validated model of the central clock~\cite{hannay2019macroscopic} through the integration of the peripheral clocks. In contrast to the single--population model, our framework provides a means to explore how competing zeitgeber inputs may be integrated across multiple tissues, and to model how the ability to remain internally synchronized in the face of changing inputs depends on factors that may change with age.  The model accounts for experimentally validated observations and suggests further hypotheses to be explored in the future. 

Here we use our model to examine the effect of three age-associated changes in the re-entrainment process. First, we find that an attenuated sympathetic pathway, represented by a decreased coupling strength between clocks in the model, attributes to disorganization in the peripheral clocks, slows down the rate of re-entrainment, and preserves (even slightly exaggerates) the asymmetry of re-entrainment. Second, our model suggests that the age-related effect of light sensitivity can lead to a loss of synchrony in the central clock and a smaller phase-advance shift. Thrid, we demonstrate that the effect of an age-related decline in coupling between cells within the central clock only explains a dampened rhythm in the brain but, surprisingly, it does not affect the re-entrainment process. In addition to investigating age-induced changes in physiology, we also show that a simple intervention---having a single meal of larger volume in the early morning for three days---can accelerate both phase delay and phase advance processes from 7 and 9 days to 4 days. 

Our model can be extended in several directions to provide further insight into the mammalian circadian system.
First, our numerical simulation assumes phase response curves to different stimuli have the same shape, which needs to be validated in future experimental studies. Different shapes of phase response curves might affect the resulting dynamics, but our framework can be easily adapted, as explained in the Method section. 
Second, we assume the external stimuli to be uniform in our analysis (regular meals, steady light). However, humans almost never experience such conditions in real life. Future studies should explore the effect of fluctuations in external stimuli, especially their effect on peripheral clocks. Addressing this question will allow us to better understand how conflicting stimuli affect circadian clocks and propose behavioral interventions. 
Third, the reduced model assumes all-to-all and instantaneous coupling for simplicity. This can be justified when the coupling is mainly achieved through neurotransmitters, whose diffusion is fast compared to the 24-hr period~\cite{shirakawa2000synchronization,shirakawa2001multiple}. However, both central and peripheral clocks can have a nontrivial network topology underlying their cellular communication. In addition, aging may also change the network topology. Therefore, future studies should address how different connection patterns affect the re-entrainment process. 
Lastly, our study demonstrates that adjusting the timing of food stimuli can be a potential avenue for correcting circadian misalignment, which calls for techniques to optimize meal schedules during an intervention.

\bibliographystyle{naturemag}
\bibliography{main}

\clearpage

\newcommand\SupplementaryMaterials{%
  \xdef\presupfigures{\arabic{figure}}
  \xdef\presupsections{\arabic{section}}
  \renewcommand{\figurename}{Supplementary Figure}
  \renewcommand\thefigure{S\fpeval{\arabic{figure}-\presupfigures}}
  \renewcommand\thesection{S\fpeval{\arabic{section}-\presupsections}}
}

\SupplementaryMaterials

\clearpage
\setcounter{page}{1}
\renewcommand{\thepage}{S\arabic{page}}

\begin{center}
{\Large\bf Supplementary Information}\\[3mm]
{\large{A minimal model of peripheral clocks reveals differential circadian re-entrainment in aging}}\\[1pt]
\end{center}

\begin{figure}[h]
    \centering
    \includegraphics[width = \textwidth]{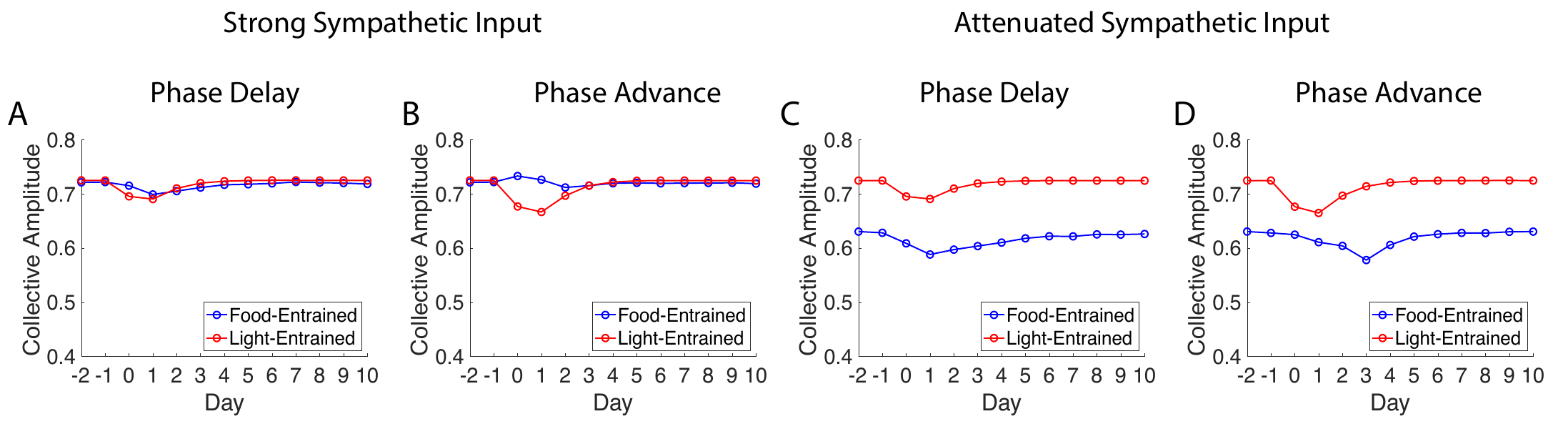}
    \caption{A decline in the input from the sympathetic pathway: Collective amplitude of the central clock and the peripheral clock, corresponding to \cref{fig2}}
    \label{fig:sup1}
\end{figure}

\begin{figure}[h]
    \centering
    \includegraphics[width = .6\textwidth]{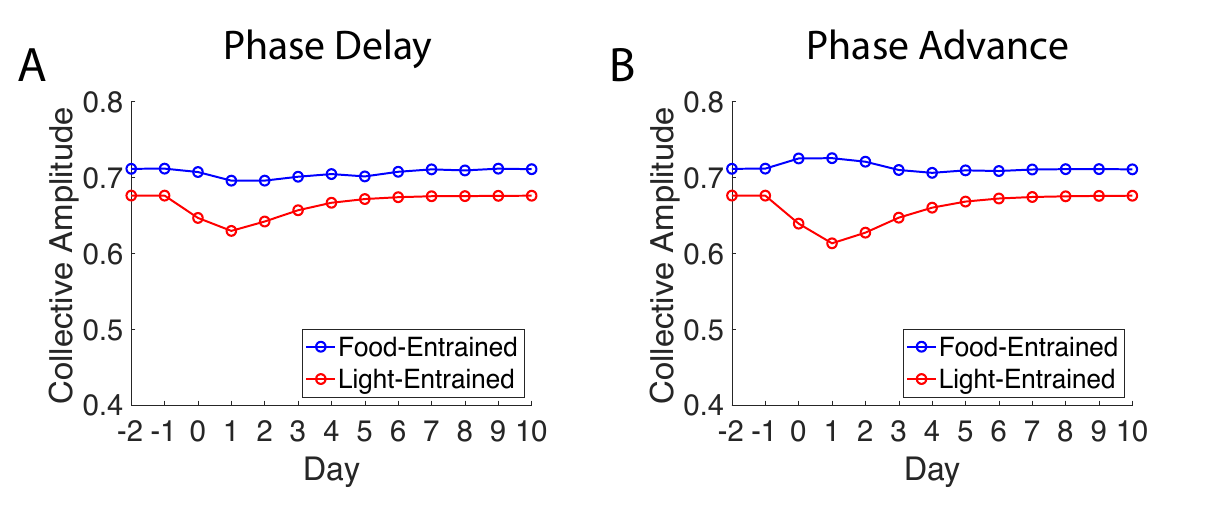}
    \caption{Reduced sensitivity to light: Collective amplitude of the central clock and the peripheral clock, corresponding to \cref{fig3}}
    \label{fig:sup2}
\end{figure}

\begin{figure}[h]
    \centering
    \includegraphics[width = .6\textwidth]{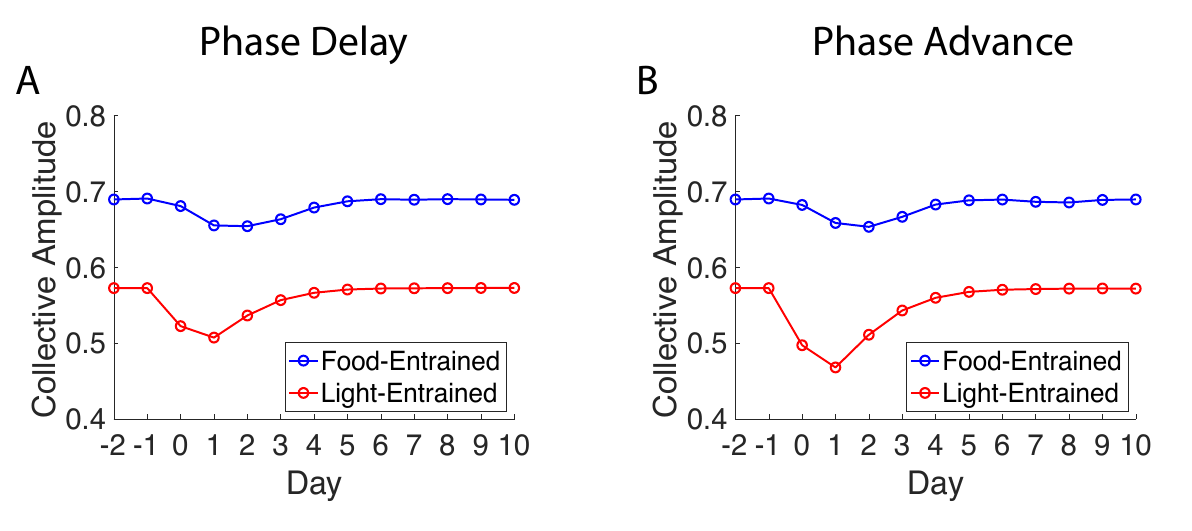}
    \caption{Reduced coupling strength in the SCN: Collective amplitude of the central clock and the peripheral clock, corresponding to \cref{fig4}}
    \label{fig:sup3}
\end{figure}

\begin{figure}[h]
    \centering
    \includegraphics[width = .6\textwidth]{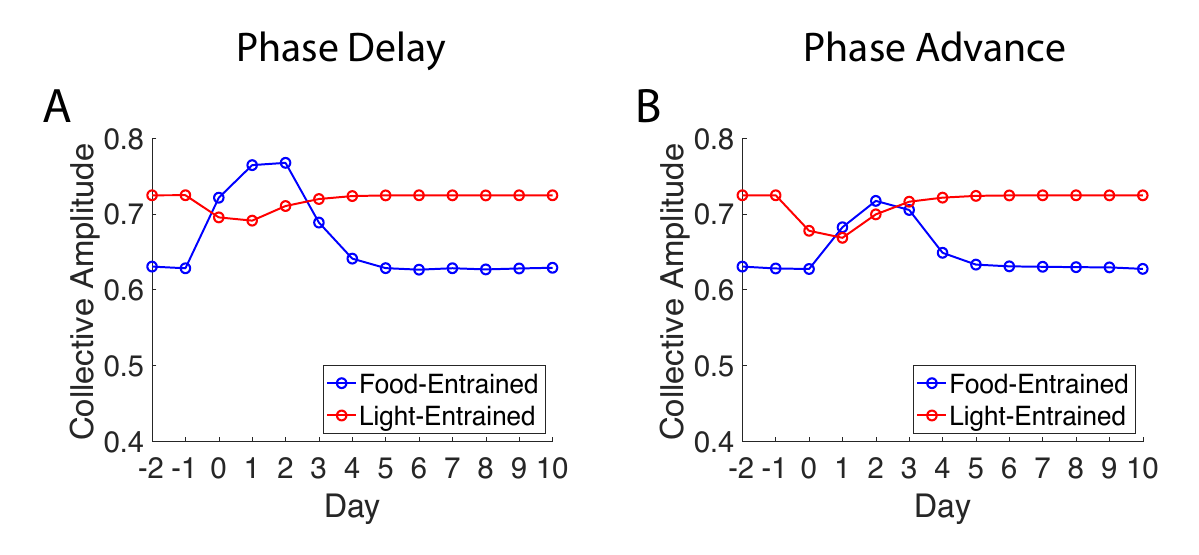}
    \caption{Meal timing as a possible rescue: Collective amplitude of the central clock and the peripheral clock, corresponding to \cref{fig5}}
    \label{fig:sup4}
\end{figure}

\end{document}